\documentclass[prl,twocolumn, showpacs,amsmath,amssymb]{revtex4}

\usepackage{graphicx}
\usepackage{amssymb}
\usepackage{amsmath}
\usepackage{enumerate}

\begin{document}

\title{Undulation instability of epithelial tissues}

\author{Markus Basan$^{1,2,3}$, Jean-Fran\c{c}ois Joanny$^{1,2,3}$, Jacques Prost$^{1,2,3,4}$ and Thomas Risler$^{1,2,3}$}
\affiliation{$^1$Institut Curie, Centre de Recherche, UMR 168, 26 rue d'Ulm, F-75005, Paris, France}
\affiliation{$^2$UPMC Univ Paris 06, UMR 168, F-75005, Paris, France}
\affiliation{$^3$CNRS, UMR 168, F-75005, Paris, France}
\affiliation{$^4$ESPCI ParisTech, F-75005, Paris, France}

\newpage

\begin{abstract}
Treating the epithelium as an incompressible fluid adjacent to a viscoelastic stroma, we find a novel hydrodynamic instability that leads to the formation of protrusions of the epithelium into the stroma. This instability is a candidate for epithelial fingering observed {\it in vivo}. It occurs for sufficiently large viscosity, cell-division rate and thickness of the dividing region in the epithelium. Our work provides physical insight into a potential mechanism by which interfaces between epithelia and stromas undulate, and potentially by which tissue dysplasia leads to cancerous invasion. 
\end{abstract}

\maketitle

Interfaces between epithelial tissues and stromas often present different degrees of undulations. In pre-cancerous abnormalities of epithelial tissues---called dysplasia---such undulations are often especially pronounced and can evolve into long fingers that extend into the stroma~\cite{tavassoli2003pag}. In a stratified epithelium, an important indicator of tissue dysplasia is the thickness of the region in which cells divide. While in healthy epithelia only the cells directly at the basement membrane divide, cell division in dysplastic tissues takes place in a larger domain, and in severe cases throughout the entire epithelium.

\begin{figure}[hb]
\begin{center}
\scalebox{0.43}{
\includegraphics{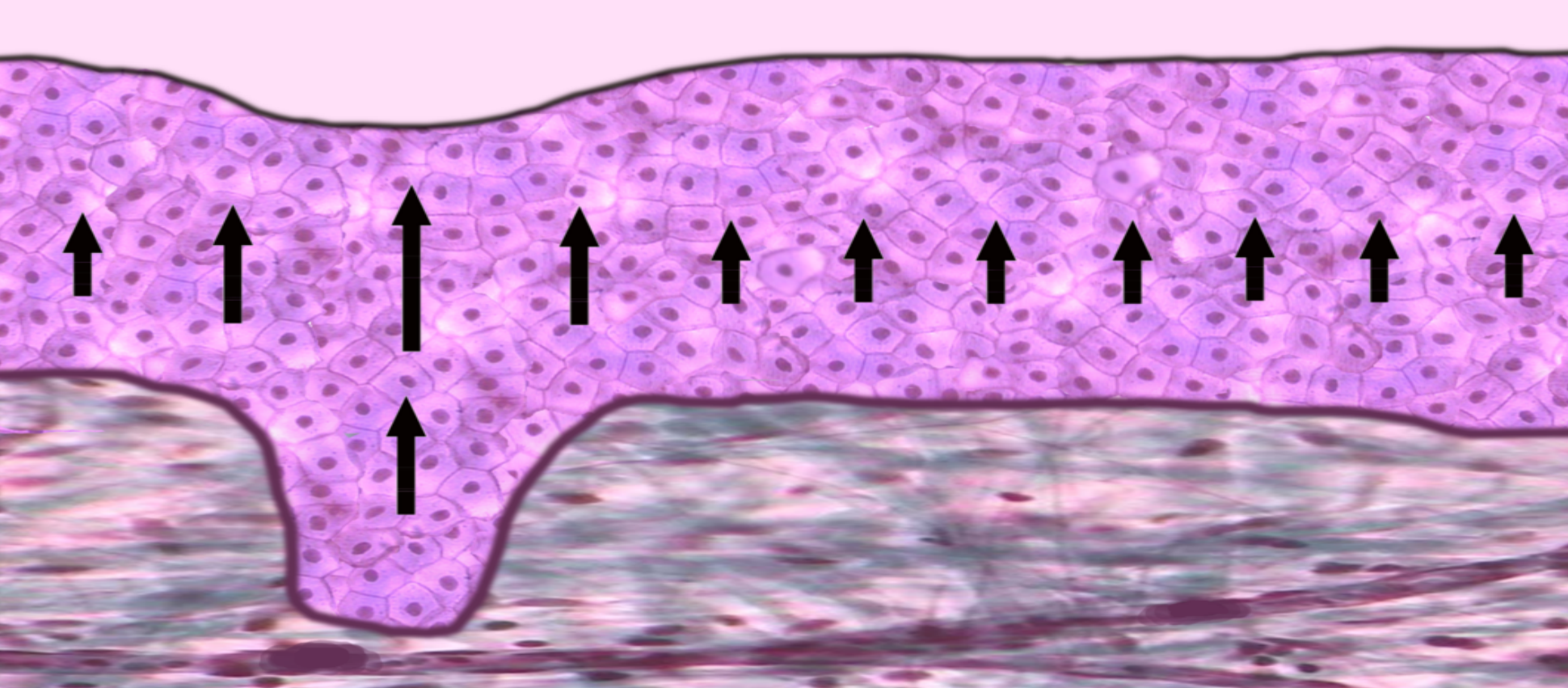}
}
\end{center}
\caption[fig1]
{\label{fig_drawing}
Schematic representation of a stratified epithelium sitting on top of its underlying stroma. The arrows represent the qualitative profile of the cell-velocity field driven by cell division. The epithelium extends a fingering protrusion into the stroma, driven by viscous shear stress.}
\end{figure}

The instability of monolayered epithelia has been modeled as the result of a buckling phenomenon~\cite{drasdo2000bio}. Other studies have used the framework of nonlinear elasticity to describe the instabilities in growing tissues~\cite{benamar}. As motivated in earlier work~\cite{basan09} and shown experimentally~\cite{steinberg,marmottant2009role}, tissues behave as viscous liquids on long time scales. This is illustrated for example by the existence of surface tension at tissue boundaries~\cite{foty1996ste,lecuit2007csm,PhysRevLett.104.218101,schotz2008qdt}. Theoretically, viscous descriptions have already been applied in other contexts of tissue growth~\cite{bittig}. Here, we propose that the fingering of a stratified epithelium originates from viscous friction effects driven by cell division. We treat the epithelium as a viscous fluid lying on top of a viscoelastic stroma (Fig.~\ref{fig_drawing}).
As the epithelium consists mostly of cells, the stroma is made of a network of collagen and elastin fibers, constantly remodeled by fibroblasts present at low densities~\cite{alberts}. On short time scales, this network responds elastically to deformations, but its constant remodeling by fibroblasts allows the tissue to flow on long time scales. A qualitative understanding of the full viscoelastic picture can be gained by interpolating between the results of the elastic and viscous regimes. In this letter, we present these two limit cases.

For the sake of simplicity, the epithelium and the stroma are each considered incompressible. In this case, the continuity equation for the epithelium reads $\partial_\alpha v_\alpha = k_{\rm p}$, where $k_{\rm p}$ is the global production rate of cells, taking into account cell division and apoptosis. The associated constitutive relation is that of an incompressible fluid with shear viscosity $\eta$ \footnote{We assume a ratio of $2/3$ between bulk and shear viscosities.}: $\sigma'_{\alpha \beta} = \eta(\partial_\alpha v_\beta + \partial_\beta v_\alpha)$. Here the total stress tensor $\sigma_{\alpha \beta}$ has been split into a dynamic part $\sigma'_{\alpha \beta}$ and a velocity-independent part $-p_{\rm e}\delta_{\alpha \beta}$, where $p_{\rm e}$ is the tissue pressure. The system of equations describing the epithelium is completed by the force-balance condition $\partial_\alpha \sigma_{\alpha \beta}=0$, which leads to:
\begin{eqnarray}\label{EqEpithelium}
\eta\partial_\alpha \partial_\alpha v_\beta + \eta\partial_{\beta}k_{\rm p} -  \partial_\beta p_{\rm e} &=& 0.
\end{eqnarray}
Similarly, for an elastic stroma, we obtain:
\begin{eqnarray}\label{elastic}
\mu\,\partial_\alpha \partial_\alpha u_\beta -  \partial_\beta p_{\rm s} = 0,
\end{eqnarray}
together with $\partial_\alpha u_\alpha = 0$, where $u_\alpha$ is the displacement field, $\mu$ the shear modulus and $p_{\rm s}$ the pressure.

Boundary conditions are as follows. The stress vanishes at the apical surface of the epithelium, taking into account the Laplace pressure due to the epithelium apical surface tension~$\gamma_{\rm a}$. At the bottom of the stroma, the displacement vanishes. At the epithelium-stroma interface, the normal component of the velocity is continuous and the normal component of the displacement of the stroma is equal to the variation of the interface location. The discontinuity of the normal component of the stress is given by Laplace's law with interfacial tension $\gamma_{\rm i}$. Finally, the tangential components of the stress are continuous and equal to a finite surface-friction term with coefficient $\xi$.

The physical origin of the instability discussed in this work can be qualitatively understood  as follows. Consider a fingering protrusion of the epithelium into the stroma and assume for simplicity that cell division occurs over the entire height of the protrusion (Fig.~\ref{fig_drawing}). The dividing cells create a flow in the epithelium. Since the cells above the finger have more dividing layers underneath them than their neighbors, they flow toward the apical surface faster than the cells in the adjacent regions. This results in a shear flow of cells within the epithelium. The associated shear stress builds up pressure at the bottom of the finger, favoring the development of the protrusion.

Let us now discuss the solution of our model for the flat, unperturbed epithelium-stroma interface. Here and in the following, we make the assumption that, due to the lack of nutrients and growth factors away from the stroma, the overall cell production decreases exponentially over a length scale $l$ with increasing distance $\Delta z$ from the epithelium-stroma interface: $k_{\rm p} = k \exp(- \Delta z/l)-k_0$ \footnote{Note that we do not expect our results to crucially depend on the detailed form of the cell-production function.}. When the interface is flat, the cell velocity and pressure in the epithelium read:
\begin{eqnarray}
v_{z}^0 &=& k l \left(1-\exp\left(- \frac{z - L}{l} \right)\right)-k_0 (z-L), \label{unpertvel}\\
p_{\rm e}^0 &=& 2 \eta \left( k \exp\left(- \frac{z - L}{l} \right) - k_0 \right),
\end{eqnarray}
where the origin of the $z$ coordinate is at the bottom of the stroma, and $L$ is the stroma thickness. The height of the epithelium $H$ is determined from the condition that the cell velocity vanishes at its apical surface. Together with Eq.~(\ref{unpertvel}), this condition reads $k_0 =  k l/H (1-\exp( - H/l))$. The deformation of the stroma vanishes everywhere \mbox{($u_{\alpha}^0=0$)}. 

We now address the question of the stability of the system under a small perturbation. Since we do not expect the origin of the instability to depend on dimensionality, we consider the case of a system translationally invariant in the $y$-direction, with a perturbation of the epithelium-stroma interface of the form $\delta h(x,t) = \delta h_0 \exp(\omega t + {\rm i} q x)$. In the linearized system of equations, the solutions for the perturbations all take this form. Eq.~(\ref{EqEpithelium}) then reads:
\begin{equation}\label{tissue}
\eta \partial_{\alpha} \partial_{\alpha} \delta v_{\beta} + \eta \partial_{\beta} \partial_{\alpha} \delta v_{\alpha} - \partial_{\beta} \delta p_{\rm e} = 0,
\end{equation}
together with the continuity equation $\partial_{\alpha} \delta v_{\alpha} = k_{\rm p}(z-\delta h) - k_{\rm p}(z) = -\frac{\partial k_{\rm p}}{\partial z}\delta h$. The bulk equations for the stroma keep their previous forms.

\begin{figure}[ht]
\begin{center}
\scalebox{0.93}{
\includegraphics{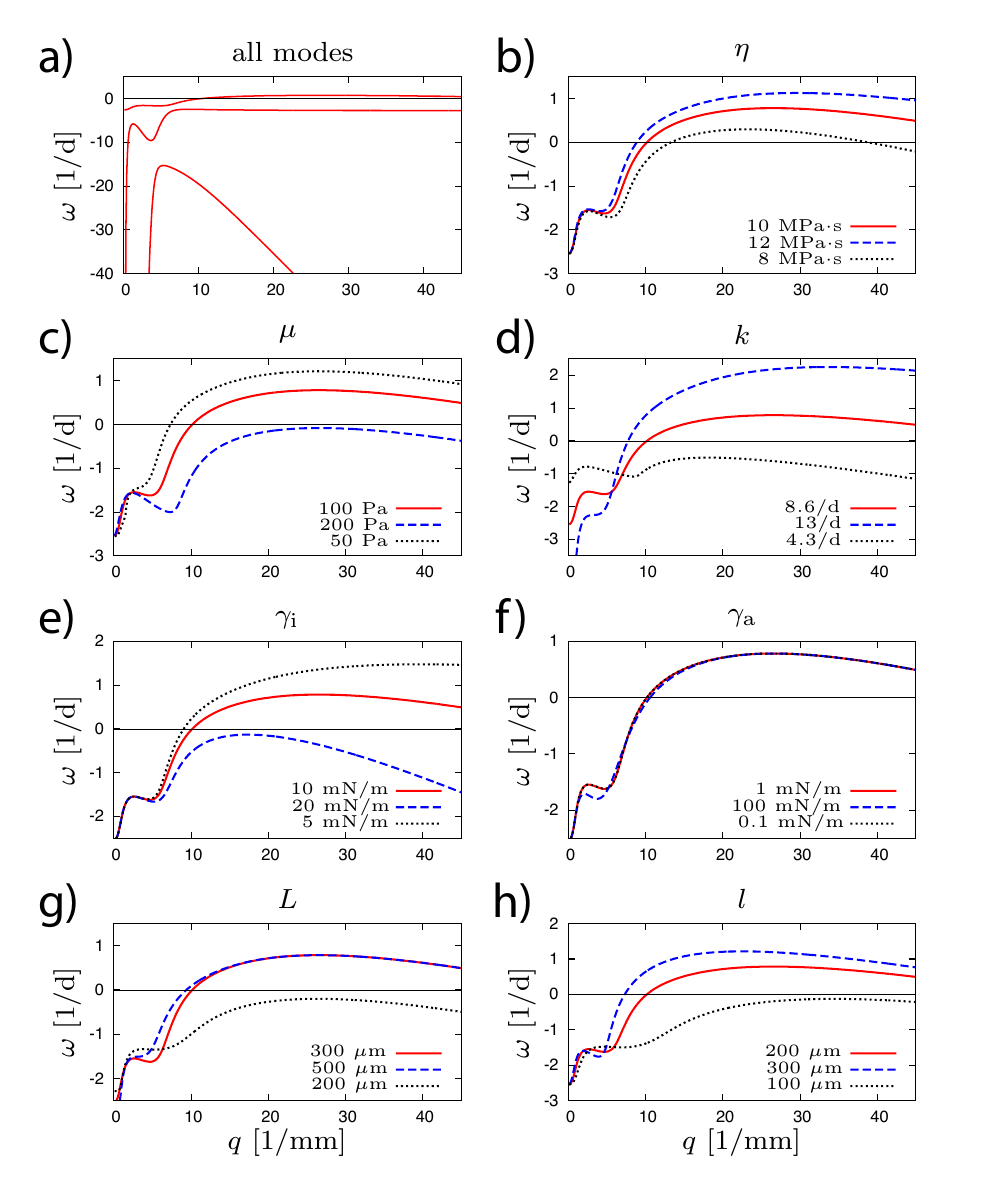}
}
\end{center}
\caption[fig2]
{\label{fig_elastic}Relaxation modes $\omega$ as a function of the wavenumber $q$ for an elastic stroma.
(a) The three relaxation modes are plotted for the following parameters: $\eta=10$~MPa$\cdot$s, $\mu=100$~Pa, $\gamma_{\rm i}=10$~mN$\cdot$m$^{-1}$, $\gamma_{\rm a}=1$~mN$\cdot$m$^{-1}$ (all estimated from~\cite{fung1993biomechanics}), $k=8.6$~d$^{-1}$ (see e.g.~\cite{weinberg2007bc}), $\xi=10^{\rm 10}$~Pa$\cdot$s$\cdot$m$^{-1}$ (estimated from~\cite{oliver1999separation}), $H=L=300$~$\mu$m and $l=200$~$\mu$m (estimated from~\cite{tavassoli2003pag}).
(b) to (h) In each panel, the most unstable mode is investigated while one parameter is varied as compared with panel (a). The varied parameter is indicated at the top of each panel, and its different values directly on each graph. Plots are coded both in color as well as line styles.
} 
\end{figure}

Stress balance at the apical surface of the epithelium reads:
\begin{eqnarray} 
\eta(\partial_x \delta v_z + \partial_{z} \delta v_x) +{\rm  i} 2 \eta (\partial_z v_z^0) q \delta H &=& 0,\label{EqTop1}\\
2 \eta \partial_z \delta v_z - \delta p_{\rm e} + \gamma_{\rm a} q^2 \delta H &=& 0.\label{EqTop2}
\end{eqnarray} 
The perturbation $\delta H$ of the apical surface is determined by the boundary condition $v_z|_{H+L+\delta H}=\omega \delta H$, which takes the form $[k \exp(-H/l) - k_0] \delta H + \delta v_z = \omega \delta H$ to linear order. Stress balance at the epithelium-stroma interface reads: 
\begin{eqnarray}\label{EqContStressInterface}
2 \eta \partial_{z} \delta v_{z} - \delta p_{\rm e} &=& 2 \mu \partial_{z} \delta u_z - \delta p_{\rm s} + \gamma_{\rm i} q^2 \delta h,\nonumber\\
\xi (\delta v_x - \omega \delta u_x) &=& \eta (\partial_x \delta v_z + \partial_{z} \delta v_x) + 2 {\rm i} \eta (\partial_z v_z^0) q \delta h,\nonumber\\
&=& \mu ( \partial_x \delta u_z + \partial_z \delta u_x).
\end{eqnarray}
Also at this interface, continuity of velocity and displacement yields $(k-k_0) \delta h + \delta v_{z} = \omega \delta h$ and $\delta u_z = \delta h$. Finally, the displacement vanishes at the bottom of the stroma: $\delta u_{\alpha} |_{z=0} = 0$. The growth rate $\omega$ is obtained by imposing the existence of a non-trivial solution to this set of linear equations. From this condition, we obtain three relaxation modes for the system (Fig.~\ref{fig_elastic}).

\begin{figure}[ht]
\begin{center}
\scalebox{0.93}{
\includegraphics{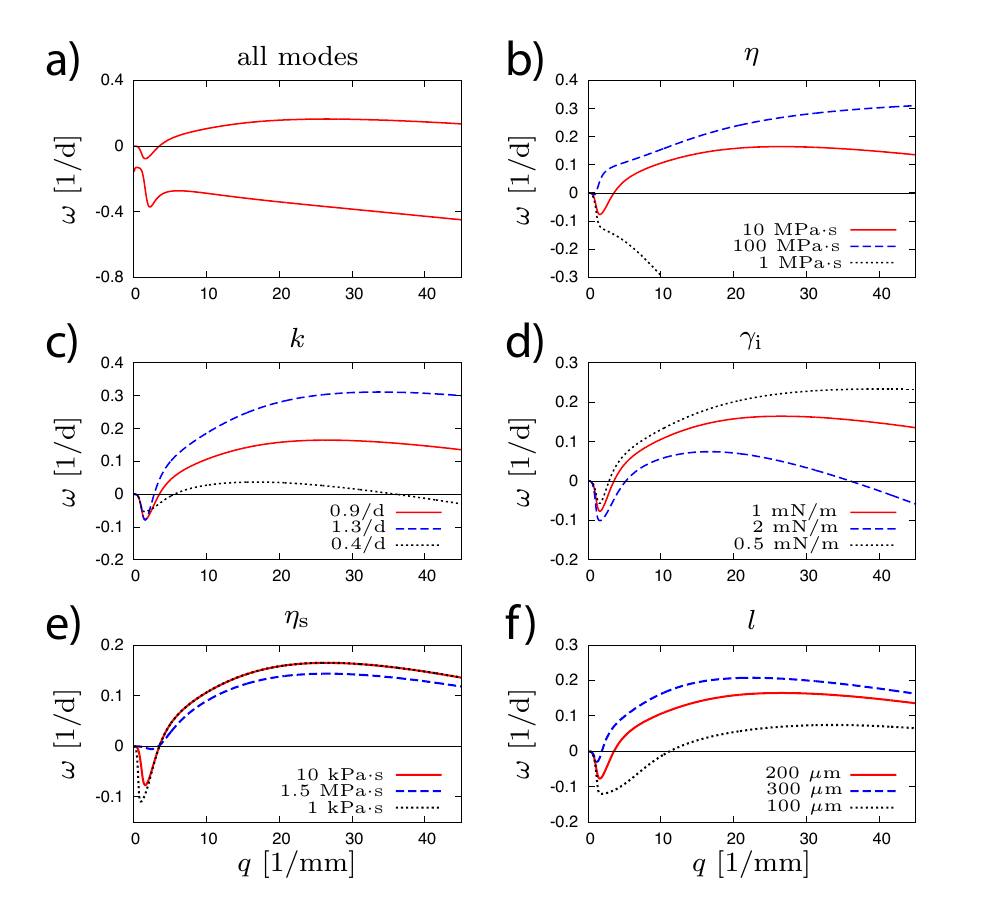}
}
\end{center}
\caption[fig3]
{\label{fig_viscous}Similar plots as those presented in Fig.~\ref{fig_elastic} and with the same conventions, but in the case of a viscous stroma. The same default parameters are used, except for $\eta_{\rm s}=10$~kPa$\cdot$s (instead of $\mu$), $\gamma_{\rm i}=1$~mN$\cdot$m$^{-1}$ and $k=0.9$~d$^{-1}$.
} 
\end{figure}

In the case where the stroma is treated as a viscous fluid, the previous equations need to be modified by replacing the displacement $u_\alpha$ by a velocity ($v_\alpha^{\rm s}$) and the shear modulus $\mu$ by a viscosity ($\eta_{\rm s}$).
In addition, the following boundary conditions are altered:
the friction term in Eq. (\ref{EqContStressInterface}) and the condition $\delta u_z = \delta h$ at the epithelium-stroma interface are replaced by $\xi (\delta v_x - \delta v^{\rm s}_x)$ and $\delta v^{\rm s}_z = \omega \delta h$, respectively. This results into two relaxation modes (Fig.~\ref{fig_viscous}).

The number of modes that we get can be understood as follows. For the elastic stroma, the set of boundary conditions generates three modes because it contains the inverse relaxation rate $\omega$ three times: in the velocity-continuity conditions at both interfaces and in the tangential stress-balance equation at the epithelium-stroma interface. In the case of a fluid stroma, we loose the mode associated with the latter equality. 

It is instructive to look at the analytic expansions of these different modes in the limit of large wave numbers~$q$. In this regime, the modes associated with respectively 
the epithelium-stroma interface and the apical surface decouple, since their characteristic decay lengths are of the order $q^{-1}$, which is much smaller than $H$. For an elastic stroma, their expansions to constant order read:
\begin{eqnarray}\label{omegaLargeqElastic}
\omega^{\rm el}_1 &\simeq& - \frac{\gamma_{\rm i}}{2 \eta}\,q - \frac{\mu}{\eta} + k - k_0,\nonumber\\
\omega^{\rm el}_2 &\simeq& - \frac{\gamma_{\rm a}}{2 \eta}\,q + k\,\textrm{e}^{-H/l} - k_0,\nonumber\\
\omega^{\rm el}_3 &\simeq& - 2 \frac{\mu}{\xi}\,q - \frac{\mu}{\eta}.
\end{eqnarray}
Among these expressions, only the one related to $\omega^{\rm el}_1$ can be positive, indicative of an unstable mode. It results at the epithelium-stroma interface from a balance of the stabilizing surface tension and stroma resistance to deformations on the one hand, and the overall positive cell-production on the other hand. This expression gives a necessary condition for the existence of an unstable regime ($\eta (k-k_0) \gtrsim  \mu$). The condition $\omega^{\rm el}_1=0$ also yields a leading-order expression for the upper crossover wavenumber from the unstable to the stable regime, provided that this crossover occurs in the large-$q$ domain. The expression for $\omega^{\rm el}_{2}$ results from a balance of surface tension and cell production at the apical surface, and the one for $\omega^{\rm el}_3$ from a balance of tangential stress and surface friction at the epithelium-stroma interface. Both expressions correspond to modes that are always stable in their region of validity.

In the case of a viscous stroma, the potentially unstable mode reads:
\begin{eqnarray}\label{omegaLargeqViscous}
\omega^{\rm v}_1 &\simeq& - \frac{\gamma_{\rm i}}{2 \left(\eta+\eta_{\rm s}\right)}\,q + \frac{\eta}{\eta+\eta_{\rm s}}\,(k - k_0).
\end{eqnarray}
The second mode has an identical expansion to that of the elastic case, and the third mode is lost.

Similar expansions can be obtained in the small-$q$ regime, but the expressions to next-to-leading order are complicated and mix the different physical origins described above. In the case of an elastic stroma, two of the three relaxation rates diverge to minus infinity, indicative of the elastic resistance of the stroma to a uniform compression. To leading order, they read:
\begin{eqnarray}\label{omegaSmallqElastic}
\tilde{\omega}^{\rm el}_1 &\simeq& - \frac{\mu}{4 \eta}\frac{1}{H L q^2},\\
\tilde{\omega}^{\rm el}_2 &\simeq& - \frac{36\mu}{\eta}\frac{1}{H^3 L^3 q^6}.
\end{eqnarray}
The third mode however has a finite small-$q$ limit, which reads $k\exp{(-H/l)} - k_0$. We can retrieve this expression by integrating the continuity equation at $q=0$ over the height of the perturbed epithelium and to leading order in the perturbations. In the case of a fluid stroma, one of the modes has the same finite limit, which is consistent with the argument presented above. However, the other relaxation rate approaches zero as $q^4$ rather than infinity:
\begin{eqnarray}\label{omegaSmallqViscous}
\tilde{\omega}^{\rm v}_1 &\simeq& -\frac{L^2\left[3H\gamma_{\rm a}+2L(\gamma_{\rm a}+\gamma_{\rm i})\right]}{6 \eta_{\rm s}}\,q^4.
\end{eqnarray}
Therefore, as the system is also always stable at sufficiently small $q$, the relaxation time diverges in this case. This is because the relaxation here is associated with lubrication-like viscous flows over large distances in the $x$-direction rather than elastic relaxation over short distances in the $z$-direction.

These results show that the instability always occurs at finite wave vector. In Figs.~\ref{fig_elastic} and \ref{fig_viscous}, we analyze the behavior of the most unstable mode as a function of the parameters. We see that the interface is destabilized when either the epithelium viscosity $\eta$, the cell-division rate $k$, or the thickness of the dividing region $l$ is increased, because of a higher resulting shear stress \footnote{Note that, since there is a relation between $k/k_0$ and $l/H$, in the case where $k$ (resp. $l$) is varied, $k_0$ (resp. $H$) is varied in proportion in order to keep the geometry (resp. the amount of cell production) constant.}. This is also true for the thickness $L$ of the stroma in the elastic case, since a thicker stroma resists less to a given deformation. Increasing the other parameters has a stabilizing effect. This is intuitive for the elastic shear modulus of the stroma $\mu$ in the elastic case and the stroma viscosity $\eta_s$ in the viscous case, as well as for the surface tension $\gamma_{\rm i}$ in both cases. The parameter $\gamma_{\rm a}$ in both cases as well as $L$ in the fluid case have little influence on the dispersion curves (not shown for the fluid case).

For a viscoelastic material with relaxation time $\tau$, we do not expect anything qualitatively different from the fluid or elastic cases to occur at large and intermediate wave vectors. In the small-$q$ regime, as the relaxation rate goes toward a finite negative value in the case of an elastic stroma, it vanishes when the latter is fluid. Getting the correct behavior in the generic viscoelastic case would require a complete study. As a general fact, we expect the curves presented in Fig.~\ref{fig_elastic} (resp. Fig.~\ref{fig_viscous}) to be valid when $\omega\tau\gg 1$ (resp. $\omega\tau\ll 1$).

In this work, we have shown the existence of a hydrodynamic instability of an interface between a viscous fluid with production terms and a viscoelastic material. The instability stems from the generation of viscous shear stress in the fluid due to material production. As such, this mechanism constitutes a new hydrodynamic instability that has not yet been described. We propose that this effect provides a potential mechanism for the undulations at epithelium-stroma interfaces {\it in vivo}. Our analysis might explain why such undulations are more pronounced in neoplastic tissues~\cite{tavassoli2003pag}. Indeed, tumorous epithelial cells divide faster than healthy cells and in a larger domain away from the basement membrane~\cite{weinberg2007bc}. The large-deformation regime of the instability might correspond to such fingering phenomena. It is commonly accepted that cancerous invasion requires the production of proteases that can degrade the basement membrane and remodel the extracellular matrix \cite{weinberg2007bc}. Such a digestion could decrease the interfacial tension between the tissues as well as the elastic modulus of the stroma, thereby triggering the present instability. The digestion of the extracellular matrix is thus not an alternative to the mechanism proposed here, but one of its determinants.  While proteases enhance the instability and allow the growth of protrusions to proceed deeper into the stroma, we expect the physical forces driving this process to originate from the mechanism presented here.

The undulation instability investigated in this work is potentially relevant for many biological systems in which interfaces of growing cell populations are present. For example, at interfaces between many tumors and healthy tissues, similar effects are observed \cite{tavassoli2003pag}. More generally, we expect this type of instability to occur in all sufficiently viscous fluids with source terms. It would therefore be interesting to conceive other systems that show the same type of instability, while being easier to characterize experimentally than living tissues.

\begin{acknowledgments}
We thank A.~Callan-Jones, M.~Lenz and X.~Sastre-Garau for many useful discussions.
\end{acknowledgments}

\end{document}